\documentclass[12pt]{article}
\usepackage{full page, 
amssymb, amscd, graphicx, 
}

    \title{{\bf Tensor categories and the mathematics of rational and
    logarithmic conformal field theory}}
    \author{Yi-Zhi Huang and James Lepowsky}
    \date{}

\begin{document}
    \bibliographystyle{alpha}
    \maketitle

    \newtheorem{rema}{Remark}[section]
    \newtheorem{propo}[rema]{Proposition}
    \newtheorem{theo}[rema]{Theorem}
   \newtheorem{defi}[rema]{Definition}
    \newtheorem{lemma}[rema]{Lemma}
    \newtheorem{corol}[rema]{Corollary}
     \newtheorem{exam}[rema]{Example}
\newtheorem{assum}[rema]{Assumption}
\newtheorem{conj}[rema]{Conjecture}
     \newtheorem{nota}[rema]{Notation}
        \newcommand{\ba}{\begin{array}}
        \newcommand{\ea}{\end{array}}
        \newcommand{\be}{\begin{equation}}
        \newcommand{\ee}{\end{equation}}
        \newcommand{\bea}{\begin{eqnarray}}
        \newcommand{\eea}{\end{eqnarray}}
        \newcommand{\nno}{\nonumber}
        \newcommand{\nn}{\nonumber\\}
        \newcommand{\lbar}{\bigg\vert}
        \newcommand{\p}{\partial}
        \newcommand{\dps}{\displaystyle}
        \newcommand{\bra}{\langle}
        \newcommand{\ket}{\rangle}
 \newcommand{\res}{\mbox{\rm Res}}
\newcommand{\wt}{\mbox{\rm wt}\;}
\newcommand{\swt}{\mbox{\scriptsize\rm wt}\;}
 \newcommand{\pf}{{\it Proof}\hspace{2ex}}
 \newcommand{\epf}{\hspace{2em}$\square$}
 \newcommand{\epfv}{\hspace{1em}$\square$\vspace{1em}}
        \newcommand{\ob}{{\rm ob}\,}
        \renewcommand{\hom}{{\rm Hom}}
\newcommand{\C}{\mathbb{C}}
\newcommand{\R}{\mathbb{R}}
\newcommand{\Z}{\mathbb{Z}}
\newcommand{\N}{\mathbb{N}}
\newcommand{\A}{\mathcal{A}}
\newcommand{\Y}{\mathcal{Y}}
\newcommand{\Arg}{\mbox{\rm Arg}\;}
\newcommand{\comp}{\mathrm{COMP}}
\newcommand{\lgr}{\mathrm{LGR}}

\newcommand{\dlt}[3]{#1 ^{-1}\delta \bigg( \frac{#2 #3 }{#1 }\bigg) }

\newcommand{\dlti}[3]{#1 \delta \bigg( \frac{#2 #3 }{#1 ^{-1}}\bigg) }

 \makeatletter
\newlength{\@pxlwd} \newlength{\@rulewd} \newlength{\@pxlht}
\catcode`.=\active \catcode`B=\active \catcode`:=\active \catcode`|=\active
\def\sprite#1(#2,#3)[#4,#5]{
   \edef\@sprbox{\expandafter\@cdr\string#1\@nil @box}
   \expandafter\newsavebox\csname\@sprbox\endcsname
   \edef#1{\expandafter\usebox\csname\@sprbox\endcsname}
   \expandafter\setbox\csname\@sprbox\endcsname =\hbox\bgroup
   \vbox\bgroup
  \catcode`.=\active\catcode`B=\active\catcode`:=\active\catcode`|=\active
      \@pxlwd=#4 \divide\@pxlwd by #3 \@rulewd=\@pxlwd
      \@pxlht=#5 \divide\@pxlht by #2
      \def .{\hskip \@pxlwd \ignorespaces}
      \def B{\@ifnextchar B{\advance\@rulewd by \@pxlwd}{\vrule
         height \@pxlht width \@rulewd depth 0 pt \@rulewd=\@pxlwd}}
      \def :{\hbox\bgroup\vrule height \@pxlht width 0pt depth
0pt\ignorespaces}
      \def |{\vrule height \@pxlht width 0pt depth 0pt\egroup
         \prevdepth= -1000 pt}
   }
\def\endsprite{\egroup\egroup}
\catcode`.=12 \catcode`B=11 \catcode`:=12 \catcode`|=12\relax
\makeatother

\def\hboxtr{\FormOfHboxtr} 
\sprite{\FormOfHboxtr}(25,25)[0.5 em, 1.2 ex] 

:BBBBBBBBBBBBBBBBBBBBBBBBB |
:BB......................B |
:B.B.....................B |
:B..B....................B |
:B...B...................B |
:B....B..................B |
:B.....B.................B |
:B......B................B |
:B.......B...............B |
:B........B..............B |
:B.........B.............B |
:B..........B............B |
:B...........B...........B |
:B............B..........B |
:B.............B.........B |
:B..............B........B |
:B...............B.......B |
:B................B......B |
:B.................B.....B |
:B..................B....B |
:B...................B...B |
:B....................B..B |
:B.....................B.B |
:B......................BB |
:BBBBBBBBBBBBBBBBBBBBBBBBB |

\endsprite

\def\shboxtr{\FormOfShboxtr} 
\sprite{\FormOfShboxtr}(25,25)[0.3 em, 0.72 ex] 

:BBBBBBBBBBBBBBBBBBBBBBBBB |
:BB......................B |
:B.B.....................B |
:B..B....................B |
:B...B...................B |
:B....B..................B |
:B.....B.................B |
:B......B................B |
:B.......B...............B |
:B........B..............B |
:B.........B.............B |
:B..........B............B |
:B...........B...........B |
:B............B..........B |
:B.............B.........B |
:B..............B........B |
:B...............B.......B |
:B................B......B |
:B.................B.....B |
:B..................B....B |
:B...................B...B |
:B....................B..B |
:B.....................B.B |
:B......................BB |
:BBBBBBBBBBBBBBBBBBBBBBBBB |

\endsprite


\begin{abstract}
We review the construction of braided tensor categories and modular
tensor categories from representations of vertex operator algebras,
which correspond to chiral algebras in physics.  The extensive and
general theory underlying this construction also establishes the
operator product expansion for intertwining operators, which
correspond to chiral vertex operators, and more generally, it
establishes the logarithmic operator product expansion for logarithmic
intertwining operators.  We review the main ideas in the construction
of the tensor product bifunctors and the associativity isomorphisms.
For rational and logarithmic conformal field theories, we review the
precise results that yield braided tensor categories, and in the
rational case, modular tensor categories as well.  In the case of
rational conformal field theory, we also briefly discuss the
construction of the modular tensor categories for the
Wess-Zumino-Novikov-Witten models and, especially, a recent discovery
concerning the proof of the fundamental rigidity property of the
modular tensor categories for this important special case.  In the
case of logarithmic conformal field theory, we mention suitable
categories of modules for the triplet $\mathcal{W}$-algebras as an
example of the applications of our general construction of the braided
tensor category structure.
\end{abstract}


\vspace{2em}

\renewcommand{\theequation}{\thesection.\arabic{equation}}
\renewcommand{\therema}{\thesection.\arabic{rema}}
\setcounter{equation}{0}
\setcounter{rema}{0}

\section{Introduction}

Vertex (operator) algebras, often called chiral algebras in the
physics literature, are a fundamental class of algebraic structures
whose extensive theory has been developed and used in recent years to
provide the means to illuminate and to solve many problems in a wide
variety of areas of mathematics and theoretical physics. In 1984,
Belavin, Polyakov and Zamolodchikov \cite{BPZ} formalized the relation
between the operator product expansion, chiral correlation functions
and representation theory, especially for the Virasoro algebra, and
Knizhnik and Zamolodchikov \cite{KZ} established fundamental relations
between conformal field theory and the representation theory of affine
Lie algebras. The mathematical notions of vertex algebra and of vertex
operator algebra were formulated in 1986 by Borcherds in \cite{B} and
in a variant form in 1988 by Frenkel-Lepowsky-Meurman in \cite{FLM2}.
The representation theory of vertex (operator) algebras plays deep
roles in both mathematics and physics, including in particular in the
representation theory of infinite-dimensional Lie algebras, the study
of sporadic finite simple groups, notably including the Monster, the
construction of knot invariants and $3$-manifold invariants, the
theory of $q$-series identities and fermionic formulas, and the study
of certain structures in algebraic geometry, as well as in conformal
field theory, string theory and quantum computing.

Since the introduction of string theory and conformal field theory in
physics and of the areas such as those mentioned above in mathematics,
many discoveries have been made by physicists and many others have
been made by mathematicians (and many jointly), and these have often
influenced what mathematicians study, and, respectively, what
physicists study, in remarkable and unexpected ways.  Benefiting
greatly from such interactions, a vast and (relatively) new
interdisciplinary area that cannot be considered to be purely
mathematics or purely physics has been blossoming for several decades
by now.  In a moment we will discuss more instances---the subjects of
the present paper---of this phenomenon.  Independently of its
applications to physics, the established mathematics of this new
area---mathematics created by both physicists and mathematicians---is
deep and is permanent, and it continues to enjoy rapid development.
Especially with the boundary between mathematics and physics sometimes
indistinct, it is a good idea to keep in mind that what counts as
established mathematics constists of theorems supported by
mathematical proofs (whether supplied by mathematicians or by
physicists), just as established physical theories are those that have
experimental confirmation (although this new interdisciplinary area is
clearly so deep, rich and broad that the time horizon for experimental
verification of the many aspects of the physics should be flexible!).
Mathematicians' and physicists' conjectures and unproved assumptions,
which should of course be recognized as such until they have proofs,
have played extremely influential roles in the development of the
theory, and we will be discussing some particularly interesting ones.

We turn to the theme of this paper---a review of braided and modular
tensor categories in rational and in logarithmic conformal field
theory.  Our theme indeed involves valuable interactions between
physicists' and mathematicians' ideas and work.

Tensor product operations for modules play central roles in the
representation theory of many important classical algebraic
structures, such as Lie algebras, groups (or group algebras),
commutative associative algebras, Hopf algebras, and in particular,
quantum groups.  They give new modules from known ones, but more
importantly, they provide powerful tools for studying modules.  Most
importantly, suitable categories of modules for such algebras,
equipped with tensor product bifunctors, contragredient functors,
appropriate natural isomorphisms, and related data, become rigid
symmetric or rigid braided tensor categories.  These tensor category
structures play such a fundamental role that many results in the
representation theory of such an algebraic structure and its
applications depend heavily on such tensor category structure. On the
other hand, a large part of these tensor category structures is so
easy to construct that these ubiquitous tensor category structures are
often not even explicitly mentioned.  (Rigidity for a tensor category,
which we will be mentioning often, is an abstraction of the classical
compatibility peroperties, for triple tensor products involving a
finite-dimensional vector space $X$ and its dual space $X^*$, relating
the natural maps $\C \to X^* \otimes X$ and $X^* \otimes X \to \C$.)

In 1988, motivated partly by Verlinde's conjecture \cite{V} on fusion
rules and modular transformations for rational conformal field
theories, Moore and Seiberg (\cite{MS1}, \cite{MS}) obtained a set of
polynomial equations for fusing, braiding and modular transformations
in rational conformal field theories, based on a number of explicit,
very strong, unproved assumptions, including in particular the
existence of a suitable operator product expansion for ``chiral vertex
operators'' (which correspond to intertwining operators, or more
precisely, to intertwining maps, in vertex operator algebra theory)
and the modular invariance of suitable traces of compositions of these
chiral vertex operators.

It is important to note that Moore and Seiberg mentioned a number of
issues, which turned out to be very substantial in the later
mathematical constructions reviewed below, that would arise if one
were to try to {\it prove} these strong assumptions using
representations of chiral algebras.

They observed an analogy between certain of these polynomial equations
and the coherence properties of tensor categories.  Later, Turaev
formulated a precise notion of modular tensor category in \cite{T1}
and \cite{T} and gave examples of such tensor categories from
representations of quantum groups at roots of unity, based on results
obtained by many people on quantum groups and their representations,
especially those in the pioneering work \cite{RT1} and \cite{RT2} of
Reshetikhin and Turaev on the construction of knot and $3$-manifold
invariants from representations of quantum groups.

On the other hand, on the rational conformal field theory side as
opposed to the quantum group side, (rigid and) modular tensor
categories for the Wess-Zumino-Novikov-Witten models, and more
generally, for rational conformal field theories, were then believed
to exist by both physicists and mathematicians, but no one had
constructed them at that time.  It was a deep unsolved problem to
construct such modular tensor categories.

This problem has now been solved.  The solution took many years and a
great deal of effort for mathematicians to eventually obtain a
complete construction of the desired modular tensor categories.

Among many mathematical works on the Wess-Zumino-Novikov-Witten
models, the works of Kazhdan-Lusztig \cite{KL1}--\cite{KL5} (which
handled negative-level analogues), Finkelberg \cite{F1}--\cite{F2},
Huang-Lepowsky \cite{HLaffine} and Bakalov-Kirillov \cite {BK} were
early explicit contributions toward the construction of the desired
modular tensor categories for this particular important class of
models.

However, even for the Wess-Zumino-Novikov-Witten models, the
construction was not accomplished until 2005, when the first author
completed a general construction \cite{HPNAS}, for all (suitable)
rational conformal field theories, of the desired (rigid and) modular
tensor categories that had been conjectured to exist; the papers
carrying out this work are discussed below.

See the end of Section 3 for a more detailed discussion of the
construction for the Wess-Zumino-Novikov-Witten models.

For rational conformal field theories in general, braided tensor
category structure on categories of modules for vertex operator
algebras satisfying suitable finiteness and reductivity conditions was
constructed by the authors in the papers \cite{tensorAnnounce},
\cite{tensor1}--\cite{tensor3}, \cite{tensorK} and \cite{tensor5}
together with the papers \cite{tensor4} and \cite{diff-eqn} by the
first author.  The modularity of these braided tensor
categories---that is, the properties of rigidity and of
nondegeneracy---were proved by the first author in \cite{Hrigidity} by
the use of the Moore-Seiberg equations, which the first author had
proved for suitable classes of representations of vertex operator
algebras in \cite{HVerlindeconjecture}.

Recall that in \cite{MS1}, \cite{MS}, these equations had been
obtained from the very strong unproved assumptions referred to above,
and these strong assumptions were even harder to prove than the
Moore-Seiberg equations.

In particular, modular tensor category structure on these module
categories was constructed in the papers mentioned above.  Certain of
the works entering into this construction also established the
operator product expansion for intertwining maps, or chiral vertex
operators (see \cite{tensor4} and \cite{diff-eqn}) and the modular
invariance of the space of $q$-traces of compositions of an arbitrary
number of these intertwining maps (see \cite{Hmodular}).

These constructions and results were established in particular in the
important special cases of the Wess-Zumino-Novikov-Witten models and
the minimal models.  But it is important to note that the theory
itself is general vertex operator algebra representation theory,
including substantial analytic reasoning; the representation theory of
affine Lie algebras and the representation theory of the Virasoro
algebra play virtually no role at all in the general theory, until one
verifies, after the general theory has been constructed, that the
hypotheses for applying the general theory hold for affine Lie
algebras at positive integral level and for the minimal models.

For {\it nonrational} conformal field theories, braided tensor
category structure on suitable categories of generalized modules for
vertex algebras satisfying suitable conditions was constructed in a
series of papers by the authors jointly with Zhang,
\cite{HLZ0}--\cite{HLZ8}, together with a paper by the first author
\cite{H12}.  These categories include those for the conformal field
theories associated to, for example, Lorentzian lattices (which do not
involve logorithms), as well as those for the much deeper logarithmic
conformal field theories.

In particular, suitable categories of modules for the triplet
$\mathcal{W}$-algebras, which are examples of vertex operator algebras
correponding to important logarithmic conformal field theories, are
braided tensor categories, by an application of our general
constructions and theory to this case.

In this paper, we review these constructions, along with the
underlying deep theory.  In the next section, we discuss the main
ideas in the constructions of the tensor product bifunctors and of the
associativity isomorphisms.  In Section 3, we review the precise
results that give modular tensor categories for rational conformal
field theories.  In this section we also briefly discuss the
construction of the modular tensor categories for the
Wess-Zumino-Novikov-Witten models and, especially, a recent discovery
concerning the proof of the rigidity property in this special case.
In Section 4, we review the precise results that produce braided
tensor categories for logarithmic conformal field theories, and in
this section we also discuss suitable categories of modules for the
triplet $\mathcal{W}$-algebras as an example illustrating the
application of our general theory.

We refer the reader to \cite{HLZ1} for a much more detailed
description of the mathematical theory and for a much more extensive
discussion of the relevant mathematics and physics literature.

\paragraph{Acknowledgments}
The first author is supported in part by NSF grant PHY-0901237. We
would like to thank the referees for their helpful comments and
suggestions.

\setcounter{equation}{0}
\setcounter{rema}{0}

\section{Ideas of the construction of the tensor product bifunctors and 
the associativity isomorphisms}

In the tensor category theory for vertex operator algebras, the tensor
product bifunctors are not built on the classical tensor product
bifunctor for vector spaces.  Correspondingly, the construction of the
natural associativity isomorphisms is highly nontrivial.  It plays a
deep role in the construction of braided and modular tensor category
structure.  In this section, we present the main ideas of the
construction of the tensor product bifunctors and the associativity
isomorphisms for suitable categories of modules or generalized modules
for vertex operator algebras.  Suitable classes of vertex algebras
more general than vertex operator algebras are also handled in
\cite{HLZ1}--\cite{HLZ8}.  Here we purposely suppress a large number
of important technical difficulties that had to be (and were)
addressed and resolved (see the authors' cited works, along with the
basic treatments \cite{FLM2}, \cite{FHL} and \cite{LL} for
background), so that the reader can see the flow of these ideas
without getting into details.  (In particular, many statements that
follow are intentionally oversimplified.  For instance, in the correct
mathmatical theory, formal variables as well as complex variables are
crucially needed.)  In the next two sections, we shall more
specifically discuss the cases of the material presented here
corresponding to rational conformal field theories and to logarithmic
conformal field theories, respectively.

At the end of this section we shall suggest a guide for further
reading.

The central concept underlying the constructions is the notion of
$P(z)$-intertwining map, where $z$ is a nonzero complex number and
$P(z)$ the Riemann sphere $\hat{\mathbb C}$ with one negatively
oriented puncture at $\infty$ and two positively oriented punctures at
$z$ and $0$, with local coordinates $1/w$, $w-z$ and $w$,
respectively, at these three punctures.

Let $V=\bigoplus_{n\in \Z}V_{(n)}$ be a vertex operator algebra,
$\mathbf{1} \in V_{(0)}$ its vacuum vector and $Y(\cdot,z)$ its vertex
operator map, which defines the algebra structure.  In language more
familiar to physicists, the vertex operators $Y(v, z)$ for $v\in V$
form a chiral algebra and the vector space underlying this chiral
algebra is isomorphic to $V$ under the map given by $Y(v, z)\mapsto
v=\lim_{z\to 0}Y(v, z) \mathbf{1}$.  In the present paper we will not
need to say much about the ever-present mode-expansion of vertex
operators such as $Y(v, z)$ in terms of powers of $z$, and we will not
need to specifically discuss the conformal vector (in $V_{(2)}$),
whose vertex operator gives the Virasoro algebra.  For a module $W$
for $V$, let $W'$ be the restricted (graded) dual module and let
$\langle \cdot, \cdot\rangle$ be the natural pairing between $W'$ and
$W$.  (In physics notation, the elements of $W'$ and $W$ are written
as $\langle \phi|$ and $|\psi\rangle$, respectively and the pairing
between $\langle \phi|$ and $|\psi\rangle$ is written as $\langle
\phi|\psi\rangle$.)

Let $W_1$, $W_2$ and $W_3$ be modules for $V$, and let $Y_1(\cdot,z)$,
$Y_2(\cdot,z)$ and $Y_3(\cdot,z)$ be the corresponding vertex operator
maps.  Given $v\in V$, the operators $Y_1(v,z)$, $Y_2(v,z)$ and
$Y_3(v,z)$ are the actions of the element $Y(v, z)$ of the chiral
algebra on $W_1$, $W_2$ and $W_3$, respectively.  Let $z$ be a fixed
nonzero complex number.  A $P(z)$-intertwining map of type ${W_3
\choose {W_1 W_2}}$ is a linear map
\begin{equation}
I: W_1 \otimes W_2 \to \overline{W}_3,
\end{equation}
where $\overline{W}_3$ is a natural algebraic completion of $W_3$
related to its ${\mathbb C}$-grading (typically, the full dual space
of the restricted dual of $W_3$), such that for $w_{(3)}'\in W_{3}'$,
$w_{(1)}\in W_{1}$, $w_{(2)}\in W_{2}$, $w_{(3)}\in W_{3}$ and $v\in
V$, the series
\begin{eqnarray*}
&\langle w_{(3)}', Y_{3}(v, z_1)I(w_{(1)}\otimes w_{(2)})\rangle,&\\
&\langle w_{(3)}', I(Y_1(v, z_1-z)w_{(1)}\otimes w_{(2)})\rangle,&\\
&\langle w_{(3)}', I(w_{(1)}\otimes Y_2(v, z_1)w_{(2)})\rangle&
\end{eqnarray*}
are absolutely convergent in the regions $|z_{1}|>|z|>0$,
$|z|>|z_{1}-z|>0$ and $|z|>|z_{1}|>0$, respectively, and for a
rational function $f(z_{1}, z)$ whose only possible poles are at
$z_{1}, z=0$ and $z_{1}=z$ and a loop $C_{1}$ in the complex plane
enclosing $z$ and $0$, we have
\begin{eqnarray*}
\lefteqn{\int_{C_{1}}f(z_{1}, z) 
\langle w_{(3)}', Y_3(v, z_1)I(w_{(1)}\otimes w_{(2)})\rangle dz_{1}}\nno\\
&&=\int_{C_{2}}f(z_{1}, z) 
\langle w_{(3)}', I(Y_1(v, z_1-z)w_{(1)}\otimes w_{(2)})\rangle dz_{1}\nno\\
&&\quad +\int_{C_{3}}f(z_{1}, z)
\langle w_{(3)}', I(w_{(1)}\otimes Y_2(v, z_1)w_{(2)})\rangle dz_{1},
\end{eqnarray*}
where $C_{2}$ is a loop in the complex plane enclosing
$z$ but not $0$ and $C_{3}$ is a loop enclosing $0$ but not $z$.

It was proved in \cite{tensor1}, \cite{tensor3} and \cite{HLZ3} that
for a nonzero complex number $z$, these $P(z)$-intertwining {\it maps}
are in fact the evaluations of intertwining {\it operators} (or
logarithmic intertwining operators, in the logarithmic theory) at $z$,
that is, given a $P(z)$-intertwining map $I$ and a choice of branch of
$\log z$, there exists an intertwining operator or logarithmic
intertwining operator $\mathcal{Y}$ of the same type such that
\begin{eqnarray}\label{intwmap=intwopatz}
I(w_{(1)} \otimes w_{(2)}) = \mathcal{Y}(w_{(1)},z)w_{(2)},
\end{eqnarray}
where the right-hand side is evaluated using the given branch of $\log
z$.  (An intertwining operator involves a {\it formal} variable, while
an intertwining map is based on a nonzero complex number $z$, as
above.  Intertwining maps correspond more closely to chiral operators
in conformal field theory, but both notions are essential in the
theory.)

There is a natural linear injection
\begin{equation}\label{homva}
\hom (W_1\otimes W_2, \overline{W}_3)\to
\hom (W'_3, (W_1\otimes W_2)^*).
\end{equation}
Under this injection, a map $I\in \hom (W_1\otimes W_2,
\overline{W}_3)$ amounts to a map $I': W'_3\to (W_1\otimes W_2)^*$:
\begin{equation}\label{I'}
w'_{(3)} \mapsto \langle w'_{(3)}, I(\cdot\otimes \cdot)\rangle.
\end{equation}
For a $P(z)$-intertwining map $I$, the map (\ref{I'}) intertwines two
natural $V$-actions, on $W'_3$ and on $(W_1\otimes W_2)^*$. The space
$(W_1\otimes W_2)^*$ is typically not a $V$-module, not even in any
weak sense.  The images of all the elements $w'_{(3)}\in W'_3$ under
this map satisfy certain very subtle conditions, called the
``$P(z)$-compatibility condition'' and the ``$P(z)$-local grading
restriction condition,'' as formulated in \cite{tensor3}, after
Proposition 13.7, and in \cite{HLZ4}, before Theorem 5.44.

Given a suitable category of generalized $V$-modules (as precisely
formulated in the cited works) and generalized modules $W_1$ and $W_2$
in this category, the $P(z)$-tensor product of $W_1$ and $W_2$ is then
defined to be a pair $(W_0,I_0)$, where $W_0$ is a generalized module
in the category and $I_0$ is a $P(z)$-intertwining map of type ${W_0
\choose {W_1 W_2}}$, such that for any pair $(W,I)$ with $W$ a
generalized module in the category and $I$ a $P(z)$-intertwining map
of type ${W \choose {W_1 W_2}}$, there is a unique morphism $\eta:
W_0\to W$ such that $I=\bar\eta \circ I_0$, where $\bar\eta :
\overline{W}_0 \to \overline{W}$ is the linear map naturally extending
$\eta$ to the completion.  This universal property characterizes
$(W_0, I_0)$ up to canonical isomorphism, {\it if it exists}.  The
$P(z)$-tensor product of $W_1$ and $W_2$, if it exists, is denoted by
$(W_1\boxtimes_{P(z)} W_2, \boxtimes_{P(z)})$, and the image of
$w_{(1)}\otimes w_{(2)}$ under $\boxtimes_{P(z)}$, an element of
$\overline{W_1\boxtimes_{P(z)} W_2}$, {\it not} of
$W_1\boxtimes_{P(z)} W_2$, is denoted by $w_{(1)}\boxtimes_{P(z)}
w_{(2)}$.

It is crucial to note that the tensor product operation depends on an
arbitrary nonzero complex number, and that we must allow this complex
number to vary, as we will explain.  Correspondingly, the resulting
tensor category structure will be much more than a braided tensor
category; it will be what we call a ``vertex tensor category,'' as
formalized in \cite{tensorK}.

{}From the definition and the natural map (\ref{homva}), one finds
that if the $P(z)$-tensor product of $W_1$ and $W_2$ exists, then its
contragredient module can be realized as the union of the ranges of all
the maps of the form (\ref{I'}) as $W'_3$ and $I$ vary.  Even if the
$P(z)$-tensor product of $W_1$ and $W_2$ does not exist, we denote
this union (which is always a subspace stable under a natural action
of $V$) by $W_1\hboxtr_{P(z)} W_2$.  If the tensor product does exist,
then
\begin{eqnarray}
W_1 \boxtimes_{P(z)} W_2 &=& (W_1 \hboxtr_{P(z)}
W_2)',\label{vertexhbox1}\\
W_1 \hboxtr_{P(z)} W_2 &=& (W_1 \boxtimes_{P(z)}
W_2)';\label{vertexhbox2}
\end{eqnarray}
examining (\ref{vertexhbox1}) will show the reader why the notation
$\hboxtr$ was chosen in the papers
\cite{tensorAnnounce}--\cite{tensor3} ($\boxtimes = \hboxtr \,'$!).

By Theorem 13.10 in \cite{tensor3} and Theorem 5.50 in \cite{HLZ4},
$W_1\hboxtr_{P(z)} W_2$ is equal to the subspace of $(W_{1}\otimes
W_{2})^{*}$ consisting of the elements satisfying the
$P(z)$-compatibility condition and the $P(z)$-local grading
restriction condition mentioned above.  With such a characterization
of $W_1\hboxtr_{P(z)} W_2$, algorithms for calculating
$W_1\hboxtr_{P(z)} W_2$ and $W_1 \boxtimes_{P(z)} W_2$ (if the tensor
product $W_1 \boxtimes_{P(z)} W_2$ of $W_1$ and $W_2$ exists) can be
developed.  It will be interesting to see how such algorithms are
related to the Nahm-Gaberdiel-Kausch algorithm (\cite{N}, \cite{GK})
used by physicists.  On the other hand, we would like to emphasize
that what our theory does is to establish the basic tensor category
structure that is assumed to exist in many works, including those
developing algorithms for calculating the equivalence class of the
tensor product of two objects in the category, and it is necessary to
use the formulations and results in our papers (for which algorithms
can also be developed) in order to establish the required tensor
category structure and associated properties (such as rigidity and
modularity, for instance).  This is a far different matter from
considering only the tensor product modules (or fusion products, or
fusion rules, or properties of fusion rules).

Even if the tensor product does not exist, the contragredient $(W_1
\hboxtr_{P(z)} W_2)'$ is still a generalized $V$-module and might
still be useful, but this generalized $V$-module might not be in the
category of generalized $V$-modules that we start with.  What is
important is that our goal has been to construct a tensor category,
not just a tensor product operation on the set of equivalence classes
of generalized $V$-modules in this category.  In particular, besides
tensor products of objects, we also need to construct tensor products
of morphisms as well as an array of natural isomorphisms relating more
than two modules.  The universal property and our $\hboxtr_{P(z)}$
constructions are crucial for the constructions of tensor products of
morphisms and, more deeply, of these required natural isomorphisms.

Here is a rough analogy from classical algebra: Before the advent of
category-theoretic understanding in the twentieth century, tensor
products of modules for a group or Lie algebra, etc., were often
described in terms of tensor products of basis elements, which would
form a basis of the desired tensor product module, whose structure
would then need to be shown to be basis-independent, or were often
described in terms of multiplicities of irreducible modules occurring
in a tensor product, corresponding to fusion rules---among other ways
of describing tensor products classically.  But the natural
associativity maps and coherence properties amounting to tensor
category structure were most smoothly formulated only later.  In the
representation theory of vertex operator algebras, the particular
formulations of tensor product that we work with are the formulations
needed for the constructions and proofs entering into the necessary
tensor category structure and rigidity and modularity, and so on.

In particular, in order to construct tensor category structure, we
need to construct appropriate natural associativity isomorphisms.
Assuming the existence of the relevant tensor products, we in fact
need to construct an appropriate natural isomorphism {}from $(W_1
\boxtimes_{P(z_1-z_2)} W_2)\boxtimes_{P(z_2)} W_3$ to
$W_1\boxtimes_{P(z_1)} (W_2\boxtimes_{P(z_2)} W_3)$ for complex
numbers $z_1$, $z_2$ satisfying $|z_1|>|z_2|>|z_1-z_2|>0$.  Note that
we are using two distinct nonzero complex numbers, and that certain
inequalities hold.  This situation corresponds to the fact that a
Riemann sphere with one negatively oriented puncture and three
positively oriented punctures can be seen in two different ways as the
``product'' of two Riemann spheres, each of them with one negatively
oriented puncture and two positively oriented punctures.

To construct this natural isomorphism, we first consider compositions
of certain intertwining maps.  As we have mentioned, a
$P(z)$-intertwining map $I$ of type ${W_3 \choose {W_1 W_2}}$ maps
into $\overline{W}_3$ rather than into $W_3$.  Thus the existence of
compositions of suitable intertwining maps always entails the
(absolute) convergence of certain series.  In particular, the
existence of the composition $w_{(1)}\boxtimes_{P(z_1)}
(w_{(2)}\boxtimes_{P(z_2)} w_{(3)})$ when $|z_1|>|z_2|>0$ and the
existence of the composition $(w_{(1)}\boxtimes_{P(z_1-z_2)}
w_{(2)})\boxtimes_{P(z_2)} w_{(3)}$ when $|z_2|>|z_1-z_2|>0$, for
general elements $w_{(i)}$ of $W_i$, $i=1,2,3$, require the proof of
certain convergence conditions.

Let us now assume these convergence conditions and let $z_1$, $z_2$
satisfy $|z_1|>|z_2|>|z_1-z_2|>0$. To construct the desired
associativity isomorphism {}from $(W_1 \boxtimes_{P(z_1-z_2)}
W_2)\boxtimes_{P(z_2)} W_3$ to $W_1\boxtimes_{P(z_1)}
(W_2\boxtimes_{P(z_2)} W_3)$, it is equivalent (by duality) to give a
suitable natural isomorphism {}from $W_1\hboxtr_{P(z_1)}
(W_2\boxtimes_{P(z_2)} W_3)$ to $(W_1 \boxtimes_{P(z_1-z_2)}
W_2)\hboxtr_{P(z_2)} W_3$.  Instead of constructing this isomorphism
directly, we embed both of these spaces, separately, into the single
space $(W_1\otimes W_2 \otimes W_3)^*$.

The space $(W_1\otimes W_2 \otimes W_3)^*$ carries a natural
$V$-action analogous to the contragredient of the diagonal action in
Lie algebra theory (as was also true for the action of $V$ on
$(W_1\otimes W_2)^*$ mentioned above).  Also, for nonzero and distinct
complex numbers $z_1$ and $z_2$ and four generalized $V$-modules
$W_1$, $W_2$, $W_3$ and $W_4$, we have a canonical notion of ``$P(z_1,
z_2)$-intertwining map {}from $W_1 \otimes W_2 \otimes W_3$ to
$\overline{W}_4$''. The relation between these two concepts comes
{}from the natural linear injection
\begin{eqnarray}
\hom (W_1\otimes W_2\otimes W_3, \overline{W}_4) &\to &
\hom (W'_4, (W_1\otimes W_2\otimes W_3)^*)\nno\\ F&\mapsto & F',
\end{eqnarray}
where $F': W'_4\to (W_1\otimes W_2\otimes W_3)^*$ is given by
\begin{equation}\label{F'}
\nu\mapsto \nu\circ F,
\end{equation}
which is indeed well defined.  Under this natural map, the
$P(z_1,z_2)$-intertwining maps correspond precisely to the maps {}from
$W'_4$ to $(W_1\otimes W_2\otimes W_3)^*$ that intertwine the two
natural $V$-actions on $W'_4$ and on $(W_1\otimes W_2\otimes W_3)^*$.

Now for generalized modules $W_1$, $W_2$, $W_3$, $W_4$, $M_1$, and a
$P(z_1)$-intertwining map $I_1$ and a $P(z_2)$-intertwining map $I_2$
of types ${W_4 \choose {W_1 M_1}}$ and ${M_1 \choose {W_2 W_3}}$,
respectively, it turns out that the composition $I_1\circ
(1_{W_1}\otimes I_2)$ exists and is a $P(z_1, z_2)$-intertwining map
when $|z_1|>|z_2|>0$.  Analogously, for a $P(z_2)$-intertwining map
$I^1$ and a $P(z_1-z_2)$-intertwining map $I^2$ of types ${W_4 \choose
{M_2 W_3}}$ and ${M_2 \choose {W_1 W_2}}$, respectively, where $M_2$
is also a generalized module, the composition $I^1\circ (I^2\otimes
1_{W_3})$ is a $P(z_1, z_2)$-intertwining map when
$|z_2|>|z_1-z_2|>0$.  Hence we have two maps intertwining the
$V$-actions:
\begin{eqnarray}
W'_4 &\to &(W_1\otimes W_2\otimes
W_3)^*\nno\\
\nu&\mapsto &\nu\circ F_1,
\label{iiv1}
\end{eqnarray}
where $F_1$ is the intertwining map $I_1\circ (1_{W_1}\otimes I_2)$,
and
\begin{eqnarray}
W'_4 &\to & (W_1\otimes W_2\otimes
W_3)^*\nno\\
\nu&\mapsto &\nu\circ F_2,
\label{iiv2}
\end{eqnarray}
where $F_2$ is the intertwining map $I^1\circ (I^2\circ 1_{W_3})$.

It is important to note that we can express these compositions
$I_1\circ (1_{W_1}\otimes I_2)$ and $I^1\circ (I^2\otimes 1_{W_3})$,
which involve intertwining {\it maps}, in terms of (logarithmic)
intertwining {\it operators}.  Let ${\cal Y}_1$, ${\cal Y}_2$, ${\cal
Y}^1$ and ${\cal Y}^2$ be the intertwining operators (or logarithmic
intertwining operators, in the logarithmic case) corresponding to
$I_1$, $I_2$, $I^1$ and $I^2$, respectively.  Then the compositions
$I_1\circ (1_{W_1}\otimes I_2)$ and $I^1\circ (I^2\otimes 1_{W_3})$
correspond to the ``product'' ${\cal Y}_1(\cdot, z_1){\cal Y}_2(\cdot,
z_2)\cdot$ and the ``iterate'' ${\cal Y}^1({\cal Y}^2(\cdot,
z_{1}-z_{2})\cdot, z_2)\cdot$ of (logarithmic) intertwining operators,
respectively.  (These products and iterates, which are obtained by
specializing formal variables to the indicated complex variables,
involve a branch of the $\log$ function and also certain convergence.)

The special cases in which the generalized modules $W_4$ are two
iterated tensor product modules, and the ``intermediate'' modules
$M_1$ and $M_2$ are two suitable tensor product modules, are
particularly interesting: When $W_4=W_1\boxtimes_{P(z_1)}
(W_2\boxtimes_{P(z_2)} W_3)$ and $M_1=W_2\boxtimes_{P(z_2)} W_3$, and
$I_1$ and $I_2$ are the corresponding canonical intertwining maps,
(\ref{iiv1}) gives the natural $V$-homomorphism
\begin{eqnarray}
W_1\hboxtr_{P(z_1)}(W_2 \boxtimes_{P(z_2)} W_3)&\to &
(W_1\otimes W_2\otimes W_3)^*\nno\\
\nu&\mapsto &(w_{(1)}\otimes w_{(2)}\otimes w_{(3)}\mapsto \nno\\
&&\quad \nu(w_{(1)}\boxtimes_{P(z_1)} (w_{(2)}\boxtimes_{P(z_2)}
w_{(3)}))),
\label{injva1}
\end{eqnarray}
and when $W_4=(W_1\boxtimes_{P(z_1-z_2)} W_2)\boxtimes_{P(z_2)} W_3$
and $M_2=W_1\boxtimes_{P(z_1-z_2)}W_2$, and $I^1$ and $I^2$ are the
corresponding canonical intertwining maps, (\ref{iiv2}) gives the
natural $V$-homomorphism
\begin{eqnarray}
(W_1 \boxtimes_{P(z_1-z_2)} W_2)\hboxtr_{P(z_2)} W_3&\to &
(W_1\otimes W_2\otimes
W_3)^*\nno\\
\nu&\mapsto &(w_{(1)}\otimes w_{(2)}\otimes w_{(3)}\mapsto \nno\\
&&\quad \nu((w_{(1)}\boxtimes_{P(z_1-z_2)} w_{(2)})\boxtimes_{P(z_2)}
w_{(3)})).
\label{injva2}
\end{eqnarray}

It turns out that both of these maps are injections, so that we are
embedding both of the spaces $W_1\hboxtr_{P(z_1)}(W_2
\boxtimes_{P(z_2)} W_3)$ and $(W_1 \boxtimes_{P(z_1-z_2)}
W_2)\hboxtr_{P(z_2)} W_3$ into the space $(W_1\otimes W_2 \otimes
W_3)^*$.  By giving a precise description of the ranges of these two
maps, it was proved in \cite{tensor4} and \cite{HLZ6} that under
suitable conditions, the two ranges are the same; this provided the
desired construction of the natural associativity isomorphisms.

More precisely, for any $P(z_1, z_2)$-intertwining map $F$, the image
of any $\nu\in W'_4$ under $F'$ (recall (\ref{F'})) satisfies certain
conditions, which we call the ``$P(z_1, z_2)$-compatibility
condition'' and the ``$P(z_1, z_2)$-local grading restriction
condition'' (see \cite{tensor4}, (14.26)--(14.29), and \cite{HLZ5},
before Proposition 8.17).  Hence, as special cases, the elements of
$(W_1\otimes W_2 \otimes W_3)^*$ in the ranges of either of the maps
(\ref{iiv1}) or (\ref{iiv2}), and in particular, of (\ref{injva1}) or
(\ref{injva2}), satisfy these conditions.

In addition, any $\lambda\in (W_1\otimes W_2\otimes W_3)^*$ induces
two ``evaluation'' maps $\mu^{(1)}_\lambda: W_{1}\to (W_{2}\otimes
W_{3})^{*}$ and $\mu^{(2)}_\lambda: W_{3}\to (W_{1}\otimes
W_{2})^{*}$, defined by $(\mu^{(1)}_\lambda(w_{(1)}))(w_{(2)}\otimes
w_{(3)})=\lambda(w_{(1)} \otimes w_{(2)}\otimes w_{(3)})$ and
$(\mu^{(2)}_\lambda(w_{(3)}))(w_{(1)}\otimes w_{(2)})=\lambda(w_{(1)}
\otimes w_{(2)}\otimes w_{(3)})$, respectively.  Any element $\lambda$
of the range of (\ref{iiv1}), and in particular, of (\ref{injva1}),
must satisfy the condition that the elements
$\mu^{(1)}_\lambda(w_{(1)})$ all lie, roughly speaking, in a suitable
completion of the subspace $W_2\hboxtr_{P(z_2)} W_3$ of $(W_2\otimes
W_3)^*$, and any element $\lambda$ of the range of (\ref{iiv2}), and
in particular, of (\ref{injva2}), must satisfy the condition that the
elements $\mu^{(2)}_\lambda(w_{(3)})$ all lie, again roughly speaking,
in a suitable completion of the subspace $W_1\hboxtr_{P(z_1-z_2)} W_2$
of $(W_1\otimes W_2)^*$.  These conditions are called the
``$P^{(1)}(z)$-local grading restriction condition'' and the
``$P^{(2)}(z)$-local grading restriction condition,'' respectively
(see \cite{tensor4}, (14.30)--(14.35), and \cite{HLZ6}, before Remark
9.7).

It turns out that the construction of the desired natural
associativity isomorphism follows {}from showing that the ranges of
both of (\ref{injva1}) and (\ref{injva2}) satisfy both of these
conditions.  This amounts to a certain condition that we call the
``expansion condition'' on our module category.  When a suitable
convergence condition and this expansion condition are satisfied, we
show that the desired associativity isomorphisms do exist, and that in
addition, the ``associativity of intertwining maps'' holds.  That is,
let $z_1$ and $z_2$ be complex numbers satisfying the inequalities
$|z_1|>|z_2|>|z_1-z_2|>0$.  Then for any $P(z_1)$-intertwining map
$I_1$ and $P(z_2)$-intertwining map $I_2$ of types ${W_4\choose {W_1\,
M_1}}$ and ${M_1\choose {W_2\, W_3}}$, respectively, there is a
suitable module $M_2$, and a $P(z_2)$-intertwining map $I^1$ and a
$P(z_1-z_2)$-intertwining map $I^2$ of types ${W_4\choose{M_2\, W_3}}$
and ${M_2\choose{W_1\, W_2}}$, respectively, such that
\begin{equation}\label{iiii}
\langle w'_{(4)}, I_1(w_{(1)}\otimes I_2(w_{(2)}\otimes w_{(3)}))
\rangle =
\langle w'_{(4)}, I^1(I^2(w_{(1)}\otimes w_{(2)})\otimes w_{(3)})
\rangle
\end{equation}
for $w_{(1)}\in W_1, w_{(2)}\in W_2$, $w_{(3)} \in W_3$ and
$w'_{(4)}\in W'_4$; and conversely, given $I^1$ and $I^2$ as
indicated, there exist a suitable module $M_1$ and maps $I_1$ and
$I_2$ with the indicated properties.  In terms of (logarithmic) intertwining
operators (recall the comments above), the equality (\ref{iiii}) reads
\begin{equation}\label{yyyy}
\langle w'_{(4)}, {\cal Y}_1(w_{(1)}, z_1){\cal
Y}_2(w_{(2)},z_2)w_{(3)}\rangle
=\langle w'_{(4)},{\cal Y}^1({\cal Y}^2(w_{(1)}, z_{1}-z_{2})w_{(2)},
z_2)w_{(3)})\rangle,
\end{equation}
where ${\cal Y}_1$, ${\cal Y}_2$, ${\cal Y}^1$ and ${\cal Y}^2$ are
the (logarithmic) intertwining operators corresponding to $I_1$, $I_2$, $I^1$ and
$I^2$, respectively.  (As we have been mentioning, the two sides of
(\ref{yyyy}) involve a branch of the $\log$ function and also certain
convergence.)  In this sense, the associativity asserts that the
product of two suitable intertwining maps can be written as the
iterate of two suitable intertwining maps, and vice versa.

{}From this construction of the natural associativity isomorphisms,
$(w_{(1)}\boxtimes_{P(z_1-z_2)} w_{(2)})\boxtimes_{P(z_2)} w_{(3)}$ is
mapped naturally to $w_{(1)}\boxtimes_{P(z_1)}
(w_{(2)}\boxtimes_{P(z_2)} w_{(3)})$ under the natural extension of
the corresponding associativity isomorphism (these elements in general
lying in the algebraic completions of the corresponding tensor product
modules). In fact, this property
\begin{equation}\label{associso}
(w_{(1)}\boxtimes_{P(z_1-z_2)} w_{(2)})\boxtimes_{P(z_2)} w_{(3)}
\mapsto
w_{(1)}\boxtimes_{P(z_1)}
(w_{(2)}\boxtimes_{P(z_2)} w_{(3)})
\end{equation}
for $w_{(1)}\in W_{1}$, $w_{(2)}\in W_{2}$ and $w_{(3)}\in W_{3}$
characterizes the associativity isomorphism
\begin{equation}
(W_{1}\boxtimes_{P(z_1-z_2)} W_{2})\boxtimes_{P(z_2)} W_{3}\to
W_{1}\boxtimes_{P(z_1)}
(W_{2}\boxtimes_{P(z_2)} W_{3}).
\end{equation}
The coherence property of the associativity isomorphisms, giving rise
to the desired tensor category structure,  follows {}from this fact.

This fact also highlights why the complex numbers parametrizing our
tensor product bifunctors must be allowed to vary; the indicated
complex numbers must satisfy the inequalities mentioned above.
Formula (\ref{associso}) is at the core of our notion of ``vertex
tensor category.''  When the various complex numbers are
systematically specialized to $z=1$, by a nontrivial procedure, we
obtain an actual braided tensor category.  But it is crucial to
realize that this specialization procedure loses information.  Our
vertex tensor category structure is much richer than braided tensor
category structure, and it provides the only route to the modular
tensor category structure, including the rigidity property, discussed
in the next section.  Braided tensor category structure carries only
``topological'' information, while this vertex tensor category
structure carries the full, and necessary, conformal-geometric and
analytic information.  We are doing ``conformal field theory,'' and it
was natural that the (necessarily) mathematical construction and
formulation of the desired braided tensor category structures would
have to be inherently conformal-geometric, for both rational and
logarithmic conformal field theories.

Note that (\ref{yyyy}) can be written as
\begin{equation}\label{yyyy2}
{\cal Y}_1(w_{(1)}, z_1){\cal Y}_2(w_{(2)},z_2)=
{\cal Y}^1({\cal Y}^2(w_{(1)}, z_1-z_2)w_{(2)},
z_2),
\end{equation}
with the ``generic'' vectors $w_{(3)}$ and $w'_{(4)}$ being implicit.
This (rigorous) equation amounts to the (logarithmic) operator product
expansion in the physics literature on (logarithmic) conformal field
theory; indeed, in our language, if we expand the right-hand side of
(\ref{yyyy2}) in powers of $z_1-z_2$ (as well as in nonnegative
integral powers of $\log (z_{1}-z_{2})$ in the logarithmic case), we
find that a product of intertwining maps is expressed as a convergent
expansion in powers of $z_1-z_2$ (and nonnegative integral powers of
$\log (z_{1}-z_{2})$), with coefficients that are again intertwining
maps, of the form ${\cal Y}^1(w,z_2)$.  When all three modules are the
vertex operator algebra itself, and all the intertwining operators are
the canonical vertex operator $Y(\cdot,x)$ itself, this operator
product expansion follows easily {}from the Jacobi identity axiom (see
\cite{FLM2}) in the definition of vertex operator algebra.  But for
intertwining operators (or logarithmic intertwining operators) in
general, it is a deep matter to construct the (logarithmic) operator
product expansion, that is, to prove the assertions involving
(\ref{iiii}) and (\ref{yyyy}) above.  This was accomplished in
\cite{tensor4} in the finitely reductive (``rational'') setting and
was considerably generalized in \cite{HLZ6} to the logarithmic
setting.

The construction of the operator product expansion, then, is
intimately related to the fact that in this theory, it is vertex
tensor categories rather than merely braided tensor categories that
form the central notion.  Braided tensor category structure alone is
not enough, but it indeed follows from our vertex tensor category
structure.

Readers who are interested in the full details of the general
construction of braided tensor category structure, in both the
rational and logarithmic cases, can consult \cite{HLZ0}--\cite{HLZ8}
and the references there.  As we mentioned in the introduction, the
paper \cite{HLZ1} gives a detailed description of the whole theory;
this includes both the part of the theory presented in the series
\cite{HLZ0}--\cite{HLZ8} and the part beyond the braided tensor
category structure.  For the modular tensor category structure
(including the rigidity) in the rational case, see \cite{HPNAS} and
\cite{Hrigidity}.

The papers \cite{HLZ1}--\cite{HLZ8} are written in an essentially
self-contained way, because what we call ``logarithmic tensor category
theory'' is new and the theory had to be constructed from the
foundations; the only prerequisities for the reader are the basics of
vertex operator algebra theory in the form presented in the relevant
parts of \cite{FLM2}, \cite{FHL} and \cite{LL}.

The theory as presented in this series includes, as a special case, a
self-contained treatment of the earlier, rational (non-logarithmic)
tensor category theory of \cite{tensor1}--\cite{tensor3} and
\cite{tensor4}, and it is in fact easier to understand that theory in
the context of the more general treatment in \cite{HLZ1}--\cite{HLZ8}
than from the original papers because of certain new results
(discussed in the introductory material in \cite{HLZ1}) in this recent
series.  In particular, the results on the rational theory in
\cite{HLZ1}--\cite{HLZ8} are stronger than the results in the original
papers.

The papers \cite{HLZ1}--\cite{HLZ8} do not specifically address
examples in detail, but references to the literature are given, and
whenever a logarithmic or non-logarithmic conformal-theoretic tensor
product (or fusion product) operation satisfies the universal property
of the present theory, and the hypotheses for applying the present
theory can be verified, then this theory provides braided tensor
category structure and, more strongly, vertex tensor category
structure and convergent operator product expansions, and, under the
appropriate further conditions, (rigid and) modular tensor category
structure.

\setcounter{equation}{0}
\setcounter{rema}{0}

\section{Modular tensor categories 
and rational conformal field theories}

Though the ideas described in the preceding section are natural, it
was highly nontrivial to carry them out completely.  Also, these ideas
do not work for general vertex operator algebras; certain subtle and
deep conditions are necessary.  Moreover, the proofs of the rigidity
and modularity in the case corresponding to rational conformal field
theories required further results, proved by the first author, beyond
the vertex-algebraic tensor category theory developed by the authors.
In this section, we expand on some comments above by discussing the
background and these results, for rational conformal field theories.

The vertex-algebraic study of tensor category structure on module
categories for suitable vertex operator algebras was stimulated by the
work of Moore and Seiberg, \cite{MS1} and \cite{MS}, in which, in the
study of rational conformal field theory, they obtained a set of
polynomial equations based on the deep and explicit assumption of the
existence of a suitable operator product expansion for ``chiral vertex
operators,'' which, as we have mentioned, correspond to intertwining
maps in vertex operator algebra theory, and they observed an analogy
between the theory of this set of polynomial equations and the theory
of tensor categories.  Earlier, in \cite{BPZ}, Belavin, Polyakov and
Zamolodchikov had already formalized the relation between the operator
product expansion, chiral correlation functions and representation
theory, for the Virasoro algebra in particular, and Knizhnik and
Zamolodchikov \cite{KZ} had established fundamental relations between
conformal field theory and the representation theory of affine Lie
algebras.  As we have discussed in the introductory material in
\cite{tensorK}, \cite{tensor1} and \cite{HLaffine}, such study of
conformal field theory is deeply connected with the vertex-algebraic
construction and study of tensor categories, and also with other
mathematical approaches to the construction of tensor categories in
the spirit of conformal field theory.  Concerning the latter
approaches, we would like to mention in particular the works of
Tsuchiya-Ueno-Yamada \cite{TUY}, Beilinson-Feigin-Mazur \cite{BFM},
Kazhdan-Lusztig \cite{KL1}--\cite{KL5}, Finkelberg
\cite{F1}--\cite{F2} and Bakalov-Kirillov \cite{BK}.

The operator product expansion and resulting braided tensor category
structure constructed by the theory in \cite{tensor1}, \cite{tensor2},
\cite{tensor3}, \cite{tensor4} were originally structures whose
existence was only conjectured: It was in their important study of
conformal field theory that Moore and Seiberg (\cite{MS1}, \cite{MS})
first discovered a set of polynomial equations from a suitable axiom
system for a ``rational conformal field theory.'' Inspired by a
comment of Witten, they observed an analogy between the theory of
these polynomial equations and the theory of tensor categories. The
structures given by these Moore-Seiberg equations were called
``modular tensor categories'' by I. Frenkel.  However, in the work of
Moore and Seiberg, as they commented, neither tensor product structure
nor other related structures were either formulated or constructed
mathematically. Later, Turaev formulated a precise notion of modular
tensor category in \cite{T1} and \cite{T} and gave examples of such
tensor categories from representations of quantum groups at roots of
unity, based on results obtained by many people on quantum groups and
their representations, especially those in the pioneering work
\cite{RT1} and \cite{RT2} of Reshetikhin and Turaev on the
construction of knot and $3$-manifold invariants from representations
of quantum groups. On the other hand, on the rational conformal field
theory side, this mathematical formulation of the notion of modular
tensor category led to a precise conjecture that the category
generated by the integrable highest weight modules of a fixed integral
positive integral level for an affine Lie algebra, and much more
generally, certain module categories for chiral algebras associated
with rational conformal field theories, could be endowed with modular
tensor category structure.  This conjecture was believed and even
assumed, but prematurely so, to be true by physicists and even
mathematicians, and its proof in fact took mathematicians many years
and a great deal of effort.

The general vertex-algebraic tensor category theory developed by the
authors in \cite{tensorAnnounce}, \cite{tensor1}--\cite{tensor3},
\cite{tensorK} and \cite{tensor5} and by the first author in
\cite{tensor4} and \cite{diff-eqn} gave a construction of braided
tensor category structure and, more importantly, as we have mentioned,
vertex tensor category structure, on the category of modules for a
vertex operator algebra satisfying suitable finiteness and reductivity
conditions. In \cite{Hmodular}, the first author proved the modular
invariance for compositions of an arbitrary number of intertwining
maps for a vertex operator algebra satisfying stronger finiteness and
reductivity conditions.  (Interestingly, Zhu's methods in his
pioneering work \cite{Zhu} on modular invariance unfortunately could
not be used or adapted to handle the necessary general case of
compositions of two or more intertwining maps, essentially because
intertwining operators do not satisfy a commutator formula, and so
new, analytic, ideas had to be introduced for the solution of this
problem in \cite{Hmodular}.)  Using this modular invariance (more
precisely, the modular invariance for compositions of two intertwining
maps), the first author proved the Moore-Seiberg equations for
suitable representations of vertex operator algebras in
\cite{HVerlindeconjecture}, at the same time providing in particular a
much stronger version of the Verlinde formula relating the fusion
rules, modular transformations, and braiding and fusing matrices than
had been previously considered.  Using these constructions and
results, the first author proved the following rigidity and modularity
result in \cite{Hrigidity} (see also \cite{HPNAS}; cf. \cite{LPNAS}):

\begin{theo}\label{mtc}
Let $V$ be a simple vertex operator algebra satisfying the conditions:
\begin{enumerate}

\item $V$ is of positive energy ($V_{(0)}=\C\mathbf{1}$ and
$V_{(n)}=0$ for $n<0$) and the contragredient $V'$, as a $V$-module,
is equivalent to $V$.

\item Every $\N$-gradable weak $V$-module is a direct sum of
irreducible $V$-modules.  (In fact, the results proved in \cite{H12}
imply that this condition can be weakened to the condition that every
grading-restricted generalized $V$-module is a direct sum of
irreducible $V$-modules.)

\item $V$ is $C_{2}$-cofinite (the quotient space $V/C_{2}(V)$ is
finite dimensional, where $C_{2}(V)$ is the subspace of $V$ spanned by
the elements of the form $\res_{z}z^{-2}Y(u, z)v$ for $u, v\in V$).

\end{enumerate}
Then the category of $V$-modules has a natural structure of rigid and
in fact modular tensor category.
\end{theo}

The following families of vertex operator algebras satisfy the three
conditions above and thus by Theorem \ref{mtc}, the category of
modules for each such vertex operator algebra has a natural structure
of (rigid and) modular tensor category:

\begin{enumerate}
\item The vertex operator algebras $V_L$ associated with positive
definite even lattices $L$; see \cite{B} and \cite{FLM2} for these
vertex operator algebras and see \cite{D1}, \cite{DL} and Section 12
of \cite{DLM} for the conditions needed for invoking Theorem \ref{mtc}
above.

\item The vertex operator algebras $L(k,0)$ associated with affine Lie
algebras and positive integral levels $k$; see \cite{FZ} for these
vertex operator algebras and \cite{FZ}, \cite{HLaffine} and Section 12
of \cite{DLM} for the conditions.  These structures correspond to the
Wess-Zumino-Novikov-Witten models.

\item The ``minimal series'' of vertex operator algebras associated
with the Virasoro algebra; see \cite{FZ} for these vertex operator
algebras and \cite{W}, \cite{H2} and Section 12 of \cite{DLM} for the
conditions.

\item Frenkel, Lepowsky and Meurman's moonshine module $V^{\natural}$;
see \cite{FLM1}, \cite{B} and \cite{FLM2} for this vertex operator
algebra and \cite{D2} and Section 12 of \cite{DLM} for the conditions.

\item The fixed-point vertex operator subalgebra of $V^{\natural}$
under the standard involution; see \cite{FLM1} and \cite{FLM2} for
this vertex operator algebra and \cite{D2}, \cite{H4} and Section 12
of \cite{DLM} for the conditions.

\end{enumerate}

In addition, the following family of vertex operator superalgebras
satisfies the conditions needed to apply the tensor category theory
developed by the authors in the series of papers
\cite{tensorAnnounce}, \cite{tensor1}--\cite{tensor3}, \cite{tensorK}
and \cite{tensor5}, and thus the category of modules for such a vertex
operator superalgebra has a natural structure of braided tensor
category:

\begin{enumerate}
\setcounter{enumi}{5}

\item The ``minimal series'' of vertex operator superalgebras
(suitably generalized vertex operator algebras) associated with the
Neveu-Schwarz superalgebra and also the ``unitary series'' of vertex
operator superalgebras associated with the $N=2$ superconformal
algebra; see \cite{KW} and \cite{A2} for the corresponding $N=1$ and
$N=2$ vertex operator superalgebras, respectively, and \cite{A1},
\cite{A3}, \cite{HM1} and \cite{HM2} for the conditions.

\end{enumerate}

It is also expected that they satisfy the three conditions in the
theorem above and thus it is expected that the category of modules for
such a vertex operator superalgebra has a natural structure of modular
tensor category.

In the special case of the second family of vertex operator algebras
listed above, those corresponding to the Wess-Zumino-Novikov-Witten
models, many mathematicians have believed for a long time (at least
twenty years) that these (rigid and) modular tensor categories must
have been constructed either by using the works of
Tsuchiya-Ueno-Yamada \cite{TUY}, and/or Beilinson-Feigin-Mazur
\cite{BFM} and Bakalov-Kirillov \cite{BK}, or by using the works of
Kazhdan-Lusztig \cite{KL1}--\cite{KL5} and Finkelberg
\cite{F1}--\cite{F2}.  In particular, the Verlinde formula conjectured
by Verlinde in \cite{V} would have been an easy consequence of such a
construction, had it indeed been achieved.

But unfortunately, it turns out that this belief has recently been
shown to be wrong.  First, it has now been known, and acknowledged,
for a while that, despite a statement in the book \cite{BK} of
Bakalov-Kirillov, the algebro-geometric methods in the works of
Tsuchiya-Ueno-Yamada \cite{TUY}, Beilinson-Feigin-Mazur \cite{BFM} and
Bakalov-Kirillov \cite{BK} cannot in fact be used to prove the
rigidity of such a tensor category or to identify the $S$-matrix for
such a tensor category with the modular transformation associated to
$\tau \mapsto -1/\tau$ on the space spanned by the ``characters'' of
the irreducible modules for such a vertex operator algebra; we shall
explain these interesting issues below.

Second, most recently, it has been discovered by the first author, and
graciously acknowledged by Finkelberg in \cite{F3}, that the works of
Kazhdan-Lusztig \cite{KL1}--\cite{KL5} and Finkelberg
\cite{F1}--\cite{F2} alone did not prove the rigidity of these tensor
categories and thus also did not identify these $S$-matrices; for such
a proof and for such an identification, one in fact needs results
proved using different methods, as we discuss.  In the course of his
argument in \cite{F2} that the categories based on modules for an
affine Lie algebra at positive integral levels could be embedded as
subquotients of Kazhdan-Lusztig's rigid braided tensor categories at
negative levels, Finkelberg proved that the elements of a certain
space are proportional to elements of a certain other space.  But it
was not proved in \cite{F2}, and it is not possible to use the methods
in \cite{F1} or \cite{F2} to prove, that these proportionality
constants are nonzero, so that one could not in fact conclude that
these two spaces are isomorphic, which had been the key step in
\cite{F2}.  This subtle issue reminded the first author that his proof
of rigidity in \cite{Hrigidity} also amounted to a proof that certain
proportionality constants are nonzero; the proof in \cite{Hrigidity}
needed the strong version of the Verlinde formula proved in
\cite{HVerlindeconjecture} involving fusion rules, modular
transformations and braiding and fusing matrices.

After the first author pointed out and simultaneously corrected the
error by invoking either (i) his general theorem in
\cite{HVerlindeconjecture} that had proved the Verlinde formula or, as
an alternative, (ii) his general theorems in \cite{Hrigidity} that had
established rigidity and identified the $S$-matrices, Finkelberg gave,
in \cite{F3}, still another alternative correction, using the Verlinde
formula proved by Faltings \cite{F} (for many but not all classes of
simple Lie algebras) and by Teleman \cite{Te} (for all classes of
simple Lie algebras).

However, in the cases (i) $E_6$ level 1 (that is, $k=1$), (ii) $E_7$
level 1, and (iii) $E_8$ levels 1 and 2, even the Verlinde formula
proved by Faltings and Teleman does not help because, as has always
been known, the works of Kazhdan-Lusztig and Finkelberg simply do not
apply to these excluded cases (and were never claimed to apply to
these cases).  In these cases, especially in the deep case $E_8$ level
2, the only proof of the rigidity and the only identification of the
$S$-matrices mentioned above were given by the first author in
\cite{Hrigidity}, using, as we have mentioned, (i) the general
vertex-algebraic tensor category theory constructed by the authors in
the papers \cite{tensorAnnounce}, \cite{tensor1}--\cite{tensor3},
\cite{tensorK} and \cite{tensor5} together with \cite{tensor4} and
\cite{diff-eqn} by the first author, and (ii) the general
vertex-algebraic theorems on modular invariance for compositions of
intertwining maps in \cite{Hmodular} and on the Verlinde conjecture in
\cite{HVerlindeconjecture}, by the first author.  Note that our theory
applies, in particular, to all the five classes of vertex operator
algebras mentioned above.  Most significantly, the theory is
vertex-algebraically conceptual and general (although, necessarily,
very elaborate) and does not exclude any individual cases (such as for
instance $E_8$ level 2 among the Wess-Zumino-Novikov-Witten models).
In \cite{KL1}--\cite{KL5}, some of the deep properties of the
constructed rigid braided tensor categories needed certain
representation theory for affine Lie algebras, including the
Knizhnik-Zamolodchikov equations, and followed {}from corresponding
properties of categories of quantum group modules for the rigidity in
particular, while the present theory is intrinsically
vertex-algebraic; as we mentioned above, the only role that affine Lie
algebras play is in the verification of the hypotheses for applying
the general theory.

{}From these discussions, we can see why the works of
Tsuchiya-Ueno-Yamada \cite{TUY}, Beilinson-Feigin-Mazur \cite{BFM} and
Bakalov-Kirillov \cite{BK} cannot in fact be used to prove the
rigidity or to identify the $S$-matrix. We discuss only the rigidity
here; the discussion for the identification of the $S$-matrix is
similar.  Substantial works using different ideas, methods and results
were needed in the proofs of the rigidity in both the work of
Finkelberg \cite{F1}--\cite{F3} and the work of the first author
\cite{Hrigidity}. The work \cite{F1}--\cite{F3} was based on the whole
theory of Kazhdan-Lusztig \cite{KL1}--\cite{KL5} on the equivalence of
the braided tensor categories associated to quantum groups and to
affine Lie algebras at negative levels.  The work \cite{Hrigidity}
needed the formula (4.29) in \cite{HVerlindeconjecture}, which in turn
needed the modular invariance for compositions of two intertwining
maps proved in \cite{Hmodular}.  Neither the equivalence of
Kazhdan-Lusztig nor the modular invariance of the first author for
compositions of intertwining maps can be proved using only the works
of Tsuchiya-Ueno-Yamada \cite{TUY}, Beilinson-Feigin-Mazur \cite{BFM}
and Bakalov-Kirillov \cite{BK}. The work \cite{HK} by Kong and the
first author provided another piece of evidence. From \cite{Hrigidity}
and \cite{HK}, it is easy to see that the rigidity is in fact
equivalent to the nondegeneracy of suitable bilinear forms on spaces
of intertwining operators among the objects of the category, or
equivalently, on suitable spaces of morphisms between objects of the
category.  The proof of this nondegeneracy in \cite{HK} needed a
formula ((4.9) in \cite{HVerlindeconjecture}) obtained in the process
of proving the Verlinde formula in \cite{HVerlindeconjecture}.  In
particular, the modular invariance proved in \cite{Hmodular} was
needed.  On the other hand, from the works of Tsuchiya-Ueno-Yamada
\cite{TUY}, Beilinson-Feigin-Mazur \cite{BFM} and Bakalov-Kirillov
\cite{BK}, bilinear forms on suitable spaces of morphisms can still be
constructed but the nondegeneracy of these forms cannot be proved
without additional input.  In fact, it has been known that these works
gave only the weak rigidity of the braided tensor category.

In \cite{Hfinitetensor}, it was pointed out that while the {\it
statement} of rigidity in fact involves only genus-{\it zero}
conformal field theory, the {\it proof} of rigidity in
\cite{Hrigidity} needs genus-{\it one} conformal field theory (the
modular invariance for compositions of intertwining maps in
\cite{Hrigidity}), and that correspondingly, there must be something
deep going on here.  The recent discovery that the works of
Kazhdan-Lusztig and Finkelberg also require knowledge of the Verlinde
formula in order to prove the rigidity in the case of affine Lie
algebras enhances this observation in \cite{Hfinitetensor}.

\setcounter{equation}{0}
\setcounter{rema}{0}

\section{Braided tensor categories and logarithmic conformal field theories}

The semisimplicity of the module categories needed in the preceding
section is related to another property of these modules, namely, that
each module is a direct sum of its ``weight spaces,'' which are the
eigenspaces of the familiar operator $L(0)$ coming {}from the Virasoro
algebra action on the module.  But there are important situations in
which module categories are not semisimple and in which modules are
not direct sums of their weight spaces. The tensor categories in this
case are intimately related to logarithmic conformal field theories in
physics.  In this section, we discuss the background and the results
in this case corresponding to logarithmic conformal field theories.

For the vertex operator algebras $L(k,0)$ associated with affine Lie
algebras, when the sum of $k$ and the dual Coxeter number of the
corresponding simple Lie algebra is not a nonnegative rational number,
the vertex operator algebra $L(k,0)$ is not finitely reductive, and,
working with Lie algebra theory rather than with vertex operator
algebra theory, Kazhdan and Lusztig constructed a natural braided
tensor category structure on a certain category of modules of level
$k$ for the affine Lie algebra in \cite{KL1}--\cite{KL5}.  This work
of Kazhdan-Lusztig in fact motivated the authors to develop an
analogous theory for vertex operator algebras rather than for affine
Lie algebras, as was explained in detail in the introductory material
in \cite{tensorAnnounce}, \cite{tensorK}, \cite{tensor1},
\cite{tensor2} and \cite{HLaffine}.  However, this general theory, in
its original form, did not apply to Kazhdan-Lusztig's context, because
the vertex operator algebra modules considered in
\cite{tensorAnnounce}, \cite{tensorK}, \cite{tensor1}, \cite{tensor2},
\cite{tensor3}, \cite{tensor4}, \cite{tensor5}, \cite{diff-eqn} are
assumed to be the direct sums of their weight spaces (with respect to
$L(0)$), and the non-semisimple modules considered by Kazhdan-Lusztig
are not in general the direct sums of their weight spaces.  Although
their setup, based on Lie theory, and ours, based on vertex operator
algebra theory, are very different, we expected to be able to recover
(and further extend) their results through our vertex operator
algebraic approach, which is very general, as we discussed above.
This motivated us, jointly with Zhang, in the work
\cite{HLZ0}--\cite{HLZ8}, to generalize the earlier work of the
authors by considering modules with {\it generalized} weight spaces,
and especially, intertwining operators associated with such
generalized kinds of modules.  As we discuss below, this required us
to use {\it logarithmic} intertwining operators, and we have been able
to construct braided tensor category structure, and even vertex tensor
category structure, on important module categories that are not
semisimple.  Using this theory, Zhang (\cite{Z1}, \cite{Z2}) has
indeed recovered the braided tensor category structure of
Kazhdan-Lusztig, and has also extended it to vertex tensor category
structure.  While in our theory, logarithmic structure plays a
fundamental role, logarithmic structure did not show up explicitly in
the work of Kazhdan-Lusztig.  As we mentioned above, the
Kazhdan-Lusztig work used properties of categories of quantum group
modules for the rigidity.  The work \cite{HLZ0}--\cite{HLZ8} and
\cite{Z1}, \cite{Z2} does not prove rigidity.

In \cite{Mi}, Milas introduced and studied what he called
``logarithmic modules'' and ``logarithmic intertwining operators'';
see also \cite{Mi2}.  Roughly speaking, logarithmic modules are weak
modules for a vertex operator algebra that are direct sums of
generalized eigenspaces for the operator $L(0)$.  Such weak modules
are called ``generalized modules'' in \cite{HLZ0}--\cite{HLZ8}.
Logarithmic intertwining operators are operators that depend not only
on (general) powers of a variable $z$, but also on its logarithm $\log
z$.

{}From the viewpoint of the general representation theory of vertex
operator algebras, it would be unnatural to study only semisimple
modules or only $L(0)$-semisimple modules; focusing, artificially, on
only such modules would be analogous to focusing only on semisimple
modules for general (not necessarily semisimple) finite-dimensional
Lie algebras.  And as we have pointed out, working in this generality
leads to logarithmic structure; thus the general representation theory
of vertex operator algebras requires logarithmic structure.

Logarithmic structure in conformal field theory was in fact first
introduced by physicists to describe Wess-Zumino-Novikov-Witten models
on supergroups (\cite{RoS}, \cite{SS}) and disorder phenomena
\cite{Gu}.  A great deal of progress has been made on this subject.
Our paper \cite{HLZ1} includes a discussion of the literature. Here we
would like to mention, in particular, \cite{FGST} for the modular
group representations relevant to logarithmic triplet models,
\cite{FHST}, \cite{GR} and \cite{Ra} for Verlinde-like formulas, and
\cite{GT} for analogues of pseudo-trace functions. One particularly
interesting class of logarithmic conformal field theories is the class
associated to the triplet $\mathcal{W}$-algebras $\mathcal{W}_{1,p}$
of central charge $1-6\frac{(p-1)^{2}}{p}$, $p=2,3,\dots$, which we
shall discuss.

Here is how such logarithmic structure also arises naturally in the
representation theory of vertex operator algebras: In the construction
of intertwining operator algebras, the first author proved (see
\cite{diff-eqn}) that if modules for the vertex operator algebra
satisfy a certain cofiniteness condition, then products of the usual
intertwining operators satisfy certain systems of differential
equations with regular singular points.  In addition, it was proved in
\cite{diff-eqn} that if the vertex operator algebra satisfies certain
finite reductivity conditions, then the analytic extensions of
products of the usual intertwining operators have no logarithmic
terms.  In the case when the vertex operator algebra satisfies only
the cofiniteness condition but not the finite reductivity conditions,
the products of intertwining operators still satisfy systems of
differential equations with regular singular points, but in this case,
the analytic extensions of such products of intertwining operators
might indeed have logarithmic terms.  This means that if we want to
generalize the results in \cite{tensorAnnounce},
\cite{tensorK}--\cite{tensor3}, \cite{tensor4} and \cite{diff-eqn} to
the case in which the finite reductivity properties are not always
satisfied, we have to consider intertwining operators involving
logarithmic terms.

Logarithmic structure also appears naturally in modular invariance
results for vertex operator algebras and in the genus-one parts of
conformal field theories.  In \cite{Hmodular}, for vertex operator
algebras satisfying the conditions in Theorem \ref{mtc}, by deriving
certain differential equations and using the duality properties for
intertwining operators, the first author was able to prove the modular
invariance for $q$-traces of products and iterates of intertwining
maps.  (As we mentioned above, the methods here for the hard part of
the argument had to be different from those in \cite{Zhu}.)  If the
vertex operator algebra is of positive energy and is $C_{2}$-cofinite
(see Conditions 1 and 3 in Theorem \ref{mtc}) but does not necessarily
satisfy Condition 2 in Theorem \ref{mtc}, suitable ``pseudo
$q$-traces'' (as introduced by Miyamoto in \cite{M}) of products and
iterates of intertwining operators still satisfy the same differential
equations, but now they involve the logarithm of $q$.  To generalize
the Verlinde conjecture proved in \cite{HVerlindeconjecture} and the
modular tensor category structure obtained in \cite{Hrigidity} on the
category of $V$-modules, one will need the general logarithmic modular
invariance of such pseudo $q$-traces of products and iterates of
intertwining maps.

The work \cite{HLZ0}--\cite{HLZ8} constructed a braided tensor
category structure on a suitable category of generalized modules for a
vertex operator algebra (or more generally a conformal vertex algebra
or a M\"{o}bius vertex algebra) under a number of natural assumptions.
In \cite{H12}, by verifying the assumptions in the papers
\cite{HLZ0}--\cite{HLZ8}, the first author proved the following
result:

\begin{theo}\label{log-btc}
Let $V$ be a $C_{2}$-cofinite vertex operator algebra of positive
energy.  Then the category of grading-restricted generalized
$V$-modules has a natural structure of braided tensor category.
\end{theo}

The main work in \cite{H12} was in proving that the category is closed
under the tensor product operation and that every finitely-generated
lower-bounded generalized $V$-module is grading restricted. The
following conjecture on rigidity was made in \cite{Hfinitetensor}:

\begin{conj}
Let $V$ be a simple vertex operator algebra satisfying Conditions 1
and 3 in Theorem \ref{mtc}. Then the braided tensor category given in
Theorem \ref{log-btc} is rigid.
\end{conj}

Triplet $\mathcal{W}$-algebras $\mathcal{W}_{1,p}$, mentioned above,
are a class of vertex operator algebras of central charge
$1-6\frac{(p-1)^{2}}{p}$, $p=2,3,\dots$ which have attracted a lot of
attention from physicists and mathematicians.  These algebras were
introduced by Kausch \cite{K1} and have been studied extensively by
physicists and mathematicians.  See the introduction of \cite{HLZ1}
for more references on the representation theory of these
algebras. Such a triplet $\mathcal{W}$-algebra satisfies the positive
energy condition and the $C_{2}$-cofiniteness condition: The
$C_{2}$-cofiniteness condition was proved by Abe \cite{A} in the
simplest case $p=2$ and by Carqueville-Flohr \cite{CF} and
Adamovi\'{c}-Milas \cite{AM1} in the general case.  Thus, as a
corollary of Theorem \ref{log-btc}, we have:

\begin{theo}\label{triplet}
The category of grading-restricted generalized $V$-modules for a
triplet $\mathcal{W}$-algebra $\mathcal{W}_{1,p}$ has a natrual
structure of braided tensor category.
\end{theo}

See Tsuchiya-Wood \cite{TW} for the rigidity of the braided tensor
category in Theorem \ref{triplet} (which was constructed, along with
the logarithmic operator product expansion, in
\cite{HLZ0}--\cite{HLZ8} together with \cite{H12}, \cite{A} (for
$p=2$), \cite{CF} and \cite{AM1}).

In general, the braided tensor category in Theorem \ref{log-btc} might
not be rigid.  An example of such a nonrigid braided tensor category
is given by the triplet $\mathcal{W}$-algebra $\mathcal{W}_{2, 3}$ of
central charge $0$ (see \cite{FGST2}, \cite{Ra1}, \cite{GRW} and \cite{AM2}).

In addition to these logarithmic issues, another way in which the
present work generalizes our earlier tensor category theory for module
categories for a vertex operator algebra is that we now allow the
algebras to be somewhat more general than vertex operator algebras, in
order, for example, to accommodate module categories for the vertex
algebras $V_L$ where $L$ is a nondegenerate even lattice that is not
necessarily positive definite (cf.\ \cite{B}, \cite{DL}); see
\cite{Z1}.

\bigskip

\noindent {\small \sc Department of Mathematics, Rutgers University,
Piscataway, NJ 08854}

\noindent {\em E-mail address}: yzhuang@math.rutgers.edu

\vspace{1em}

\noindent {\small \sc Department of Mathematics, Rutgers University,
Piscataway, NJ 08854}

\noindent {\em E-mail address}: lepowsky@math.rutgers.edu

\end{document}